\documentstyle[emulateapj]{article}
\submitted{accepted to the Astrophysical Journal}

\def\v{\mbox{\boldmath{$v$}}}

\begin{document}

\lefthead{
Mori \& Burkert
}

\righthead{
Gas Stripping of Dwarf Galaxies in Clusters of Galaxies
}

\title{
Gas Stripping of Dwarf Galaxies in Clusters of Galaxies
}

\author{
Masao Mori\altaffilmark{1} and
Andreas Burkert
}

\affil{
Max-Planck-Institut f\"{u}r Astronomie,
K\"{o}nigstuhl 17, D-69117 Heidelberg, Germany
}

\altaffiltext{1}{
Institute of Astronomy, University of Tokyo,
Mitaka, Tokyo 181-8588, Japan
}

\begin{abstract}
Many clusters of galaxies contain an appreciable amount of
hot gas, the intracluster medium. As a consequence, gas will
be stripped from galaxies that  move through the inter cluster
medium, if the ram pressure exceeds the internal gravitational
force. We study the interaction between the intracluster
medium and an extended gas component of dwarf galaxies
confined by a surrounding cold dark matter halo analytically
and numerically, using axisymmetric two-dimensional
hydrodynamical simulations at high resolution. The results
show that the gas within the dark matter halo is totally
stripped in a typical galactic cluster. The process of ram
pressure stripping therefore must have played an important
role during the chemo-dynamical evolution of dwarf galaxies in
galactic clusters.
Our results predict that most of the chemical evolution of 
dwarf galaxies must have occurred at high redshift, before 
the virialized cluster had formed.
\end{abstract}


\section{Introduction}
Explaining the morphology-density relation (Dressler 1980),
that is the higher early-type galaxy fraction in clusters of
galaxies in contrast to the higher late-type galaxy fraction
in the fields, remains one of the most important problems in
cosmology. This relation holds also for dwarf galaxies.
Gas-poor dwarf galaxies are strongly clustered, while gas-rich
dwarf galaxies appear to be the most weakly clustered objects
(Binggeli et al. 1987; Ferguson \& Sandage 1988; Binggeli et
al. 1990, Thuan et al. 1991).

Since many clusters of galaxies contain an appreciable amount
of hot gas, the intracluster medium (ICM), gas will be
stripped from galaxies that move through the ICM, if the ram
pressure exceeds the internal gravitational force.
There are actually many observational evidences for the ram
pressure stripping of giant galaxies in clusters of galaxies
(Irwin et al. 1987; White et al. 1991; B\"{o}hringer et al.
1995). In addition, the physics of ram pressure stripping of
giant galaxies has been studied in detail using hydrodynamic
simulations (e.g. Balsara, Livio, \& O'Dea 1994 and references
therein). Since dwarf galaxies have smaller escape velocities,
ram pressure stripping is expected to be more efficient than
for giant galaxies. However, no attempts have ever been made
to examine the amount of gas that could be stripped from
low-mass galaxies with shallow gravitational potential wells
during their passage through an ICM.

This situation motivated us to consider the ram pressure
stripping of the diffuse gas phase of dwarf galaxies
by the ICM which may be of prime importance to understand the
morphology-density relation for dwarf galaxies. We estimate
the critical condition for gas removal by the interaction
between an ICM and gaseous dwarf galaxies confined by
a surrounding cold dark matter (CDM) halo. Moreover, we
verified that condition using an axial symmetric
two-dimensional hydrodynamic code which is based on the
piecewise parabolic method (PPM) described by Colella \&
Woodward (1984).

In \S 2, prior to the main subject, we describe the model of
dwarf galaxies which is treated as a one-parameter family
defined by the core mass of the CDM halos, using the observed
scaling relation (Burkert 1995).
In \S 3, we give the analytical estimates of the critical
conditions for the gas ablation from dwarf galaxies due to ram
pressure stripping by the ICM. In \S 4, we demonstrate the
complex interaction between dwarf galaxies and the ICM using
hydrodynamical simulations and compare the simulation results
with the analytical predictions. In \S 5, we summarise the
results of this paper, and discuss observational and
theoretical implications.


\section{Model of dwarf galaxies}
The hierarchical clustering model of galaxies has been
successful in explaining the clustering pattern of galaxies
revealed by redshift surveys.
On the other hand, the hierarchical model predicts a large
number of low-mass galaxies formed at high redshifts beyond
that estimated from the observed luminosity function of
galaxies (White \& Frenk 1991; Cole et al. 1994).
Therefore, several mechanisms for suppressing and/or delaying
the formation of dwarf galaxies have been proposed in order
to remove this serious discrepancy.

Using a three-dimensional $N$-body/SPH simulation code
combined with stellar population synthesis, Mori et al. (1997)
showed that an energy feedback via supernovae and stellar
winds from massive stars keeps a large fraction of the
gas hot and suppresses the
formation of dwarf galaxies.
Moreover, they pointed out that  this energy feedback is 
a necessary mechanism to reproduce the internal structure and the
photometric quantities of the nearby dwarf galaxies
(see also Dekel \& Silk 1986; Yoshii \& Arimoto 1987;
Burkert \& Ruiz-Lapuernnte 1997; Mori, Yoshii \& Nomoto 1999).

Babul \& Rees (1992) and Efstathiou (1992) argued that the
formation epoch of dwarf galaxies is delayed until $z<1$ due
to the photoionization of the gas by the ultraviolet
background radiation at high redshifts.
The ionizing background at $z>1$ is high enough to keep the
gas in dwarf galaxy halos confined and stable, neither able
to escape, nor able to collapse
(see also Rees 1986; Ikeuchi 1986;
Katz, Weinberg \& Hernquist 1996;
Navarro \& Steinmetz 1997).

Following this scenario, we assume that a (proto-) dwarf
galaxy has an already virialized CDM halo and a large amount
of extended hot gas heated either due to a first
population of supernovae or due to the
photoionization by ultraviolet background radiation at the
formation epoch of the galaxy clusters.

\subsection{Structure of the CDM halos}
The formation of the CDM halos through hierarchical clustering
predicts that the equilibrium density profile of a CDM halo
has a central cusp (Dubinski \& Carlberg 1991; Navarro, Frenk
\& White 1996; Fukushige \& Makino 1997; Moore et al. 1998).
However, recent observed rotation curves of nearby dwarf
galaxies rule out the singular density profile of dark matter
halos (Flores \& Primack 1994; Moore 1994). Moreover, Burkert
(1995) points out that the density profile
\begin{eqnarray}
\rho_{\rm d}(r) = \frac{\rho_{\rm d0} r_0^3}
	{(r+r_0)(r^2+r_0^2)}, 
\label{rhod}
\end{eqnarray}
nicely reproduces the rotation curves of nearby dwarf
galaxies and the central density $\rho_{\rm d0}$ is
correlated with the core radius $r_0$ through a simple scaling
relation.

In this paper we assume that the density distribution of the
CDM halos is represented by Burkert's (1995) profile for a
wide range of core masses from $10^6 M_\odot$ to $10^{10}
M_\odot$. For the density distribution of the CDM halo in
equation (\ref{rhod}),the potential is found by integrating
Poisson's equation as
\begin{eqnarray}
\Phi_{\rm d}(r)&=&-\pi G \rho_{\rm d0} r_0^2
\left[
   \pi -2 \left(  1 + \frac{r_0}{r}      \right)
	  \arctan \frac{r_0}{r} \right.  \nonumber \\
 &+&    2 \left(  1 + \frac{r_0}{r}    \right)
       \ln\left(  1 + \frac{r}{r_0}    \right) \nonumber \\
 &-&      \left. \left(  1 - \frac{r_0}{r}     \right)
       \ln\left\{ 1 + \left(\frac{r}{r_0}\right)^2   \right\}
\right]. \label{pot}
\end{eqnarray}
The central potential is given by $\Phi_{\rm d}(0)=-\pi^2 G
\rho_{\rm d0} r_0^2$, where $G$ is the gravitational constant.
The mass distribution of the CDM halo is given by
\begin{eqnarray}
M_{\rm d}(r)=\pi \rho_{\rm d0} r_0^3
\left[
   -2 \arctan \frac{r}{r_0} \makebox[7.5em]{}
				\right. \nonumber \\
\left.
   +2 \ln \left( 1 + \frac{r}{r_0} \right)
	+ \ln \left\{ 1 + \left( \frac{r}{r_0}\right) ^2
				\right\}
\right]. \label{mass}
\end{eqnarray}

Using the observed scaling relation derived by Burkert (1995),
$\rho_{\rm d0}$ and $r_0$ are calculated as
\begin{eqnarray}
r_0 = 3.07 \left( \frac{M_0}{10^9 ~M_\odot}
		\right)^{\frac{3}{7}}
		~{\rm kpc}, \label{r0}
\end{eqnarray}
and
\begin{eqnarray}
\rho_{\rm d0} = 1.46 \times 10^{-24}
	\left( \frac{M_0}{10^9 ~M_\odot}
		\right)^{-\frac{2}{7}}
		~{\rm g ~cm}^{-3}, \label{rho0}
\end{eqnarray}
where the core mass $M_0$ is the total mass of the CDM halo
inside $r_0$.

The circular velocity $v_{\rm c}^2=GM_{\rm d}(r)/r$ of the CDM
halo has a maximum value
\begin{eqnarray}
v_{\rm c, max} = 48.7
	\left( \frac{M_0}{10^9 ~M_\odot}
		\right)^{\frac{2}{7}}
		~{\rm km ~s}^{-1}, \label{vmax}
\end{eqnarray}
at the radius of $r_{\rm c, max}=3.24 r_0$ (see Fig. 1).
This velocity and the radius are related by
\begin{eqnarray}
v_{\rm c, max} = 10.5
	\left( \frac{r_{\rm c, max}}{\rm kpc}
		\right)^\frac{2}{3}
		~{\rm km ~s}^{-1}. \label{v-r}
\end{eqnarray}
%
%

\subsection{Structure of the gas}
We focus on the general question of the interaction between
the intracluster medium and a hot gaseous component in dwarf
galaxies that is generated e.g.  by the supernova heating
(Dekel \& Silk 1986; Burkert \& Ruiz-Lapuente 1997;
Mori et al. 1997; Mori et al. 1999), or by the photoionization
due to the ultraviolet background radiation (Efstathiou 1992;
Babul \& Rees 1992). In this case the gas is pressure supported
justifying the assumption of an initially spherically symmetric
distribution and of axi-symmetry.

The self-gravity of the gas is neglected for simplicity and
the gas is assumed to be in hydrostatic equilibrium with the
constant temperature
\begin{eqnarray}
T&=&\frac{\mu m_{\rm p}}{3 k_{\rm B}} \frac{G M_0}{r_0},
		\label{temp} \\
 &=&3.45\times10^4\left( \frac{M_0}{10^9 ~M_\odot}
		\right)^\frac{4}{7} ~{\rm K},
\end{eqnarray}
where $\mu$ is the mean molecular weight, $m_{\rm p}$ is the
proton mass, and $k_{\rm B}$ is the Boltzmann's constant.
The density distribution of the gas is given by
\begin{eqnarray}
\rho_{\rm g}(r)=\rho_{\rm g0}
{ \left[ 
       -\frac{\mu m_{\rm p}} {k_{\rm B} T}
	\left\{\Phi_{\rm d}(r) - \Phi_{\rm d}(0) \right\}
        \right] },\label{rhog}
\end{eqnarray}
where $\rho_{\rm g0}$ is the central density of the gas. The 
mass ratio between the gas and the CDM halo within a core 
radius is defined as
\begin{eqnarray}
F = \frac{M_{\rm g0}}{M_0}, \label{F}
\end{eqnarray}
where $M_{\rm g0}$ is the total gas mass inside a core radius.
The relation between the central density of the gas and of the 
dark matter is given by the numerical solution of equation 
(\ref{F}): $\rho_{\rm g0}=1.40 F \rho_{\rm d0}$.
We assumed initially $F=0.1$ in this paper.
Consequently each dwarf galaxy is described as a one-parameter 
family by the core mass $M_0$ of the CDM halos. 

\section{Analytical model}
The stripping process is generally classified into two 
distinct types such as the instantaneous stripping phase and 
the continuous stripping phase due to Kelvin-Helmholtz 
instability. In this section, we estimate analytically the gas 
ablation affects on the evolution of a dwarf galaxy.

\subsection{Instantaneous ram pressure stripping}
We assume that the CDM halo associated with the dwarf galaxies 
moves in a homogeneous ICM with number density of 
	$n_{\rm CG}=
	\rho_{\rm CG}/(\mu m_{\rm p})\sim10^{-4}$ cm$^{-3}$ 
and relative velocity 
	$v_{\rm gal} \sim 10^3$ km s$^{-1}$ 
corresponding to the three-dimensional velocity dispersion of 
the galaxy cluster.

The complete condition for the ram pressure stripping from a 
dwarf galaxy requires that the ram pressure of the ICM 
exceeds the thermal pressure 
	$\rho_{\rm g0} k_{\rm B} T/(\mu m_{\rm p})$ at the 
center of the gravitational potential well of the CDM halo.  
Using equation (\ref{temp}), such a condition is described by
\begin{eqnarray}
\rho_{\rm CG} v_{\rm gal}^2 >
		\frac{G M_0 \rho_{\rm g0}}{3r_0}. \label{ce1}
\end{eqnarray}
Thus, if the core mass $M_0$ is smaller than the critical core 
mass
\begin{eqnarray}
M_{\rm IS} = 1.27\times10^9 \left( \frac{F}{0.1} 
		\right)^{-\frac{7}{2}} 
\left( \frac{n_{\rm CG}}{10^{-4}{\rm cm}^{-3}} 
		\right)^{\frac{7}{2}}
		\makebox[2.5em]{} \nonumber \\
\times 
\left( \frac{v_{\rm gal}}{10^{ 3}{\rm km ~s}^{-1}} 
		\right)^{7} M_\odot,
		\label{mis}
\end{eqnarray}
the gas in a dwarf galaxy is totally stripped by the ram 
pressure of the ICM.

The collision between the gas of the dwarf galaxy and the ICM 
generates a forward shock that propagates through the gas of 
the dwarf galaxy and a reverse shock that propagates through 
the ICM. The time-scale of the gas removal is roughly 
estimated by dividing the diameter of the core by the velocity 
of the forward shock $v_{\rm fs}$. 
The analysis of the one-dimensional shock problem gives 
\begin{eqnarray}
v_{\rm fs} = \frac{4}{3}
	\sqrt{\frac{\rho_{\rm CG}}{\rho_{\rm g0}}}v_{\rm gal},
\end
{eqnarray}
for the high-speed collision of two homogeneous
non-gravitating media with large density ratio ($\rho_{\rm CG}
\ll \rho_{\rm g0}$).
The time-scale of the mass removal from dwarf galaxies is
estimated by
\begin{eqnarray}
\tau_{\rm IS}=\frac{2 r_0}{v_{\rm fs}}, \makebox[17.9em]{}\\
=2.02\times10^8\left(\frac{F}{0.1}\right)^\frac{1}{2}
 \left(\frac{M_0}{10^9M_\odot}\right)^\frac{2}{7}
	\makebox[6em]{} \nonumber \\
 \times
 \left(\frac{n_{\rm CG}}{10^{-4}{\rm cm}^{-3}}
		\right)^{-\frac{1}{2}}
 \left(\frac{v_{\rm gal}}{10^{ 3}{\rm km ~s}^{-1}} \right)^{-1}
	{\rm yr}. \label{tauss}
\end{eqnarray}

Consequently, the gas is totally stripped from less-massive
dwarf galaxies ($M_0 \lesssim 10^9M_\odot$) in the typical
environment of the ICM because the ram pressure of the ICM
exceeds the gravitational force of these galaxies.
The time-scale of the gas removal is small compared with the
characteristic time-scale (several Gyr) a galaxy spends within
potentials, therefore, lose their gas easily within this
dwarf galaxies ($M_0 \gtrsim 10^9M_\odot$) the gas might be
significantly removed except around the central region of the
gravitational potential well.

\begin{deluxetable}{crrrrr}
\tablecaption{Parameters of the ICM}
\tablehead{
	\colhead{Model}         &
	\colhead{$n_{\rm CG}$}  &
	\colhead{$v_{\rm gal}$} &
	\colhead{$T_{\rm CG}$\tablenotemark{a}} &
	\colhead{$M$\tablenotemark{b}}          &
	\colhead{$\rho_{\rm CG} v_{\rm gal}^2$} \\
	\colhead{}                 &
	\colhead{cm$^{-3}$}        &
	\colhead{km~s$^{-1}$}      &
	\colhead{K}                &
	\colhead{}                 &
	\colhead{erg~cm$^{-3}$}
}
\startdata
 ($a$)  &     $1.0\times10^{-3} $ & 1000 & $1.0\times10^7$ &
	2.1 & $1.0\times10^{-11}$ \nl
 ($b$)  &      $1.0\times10^{-3}$ &  500 & $1.0\times10^7$ &
	1.1 & $2.6\times10^{-12}$ \nl
 ($c$)  &      $1.0\times10^{-4}$ & 1000 & $1.0\times10^7$ &
	2.1 & $1.0\times10^{-12}$ \nl
 ($d$)  &      $1.0\times10^{-4}$ &  500 & $1.0\times10^7$ &
	1.1 & $2.6\times10^{-13}$ \nl
 ($e$)  &      $1.0\times10^{-5}$ & 1000 & $1.0\times10^7$ &
	2.1 & $1.0\times10^{-13}$ \nl
\enddata
\tablenotetext{a}{Temperature of the ICM}
\tablenotetext{b}{Mach number of the ICM}
\end{deluxetable}

\subsection{Kelvin-Helmholtz instability}
Gas in the massive dwarf galaxies surviving the instantaneous
ram pressure stripping is subsequently removed by
Kelvin-Helmholtz instability occurring at the interface
between the gas in the dwarf galaxy and the ICM.
Murray et al.(1993) estimated the characteristic growth time
of Kelvin-Helmholtz instability for a dense cloud embedded in
a low-density background. They showed that the growth time is
comparable to the sound crossing time of the gas cloud if
gravity is negligible.
On the contrary, the presence of a gravitational field by the
surrounding CDM halo tends to suppress the instability and to
stabilize the gas against removal from the dwarf galaxy.

We suppose that the process of the instanteneous ram pressure
stripping removes the gas distributed outside a core radius
$r_0$. Moreover, a sharp discontinuity of the gas density is
established between the ICM and the galactic gas having
nearly constant density after the shock passage. The unstable
wave number of Kelvin-Helmholtz instability at the interface
for the incompressible fluid is given by
\begin{eqnarray}
k > g \frac{\rho_{\rm CG}^2 -\rho_{\rm g, avg}^2}
	   {\rho_{\rm CG}    \rho_{\rm g, avg} v_{\rm gal}^2},
		\label{ce2}
\end{eqnarray}
where $\rho_{\rm g, avg}$ is the mean gas density inside
$r_0$ and $g=GM_0/r_0^2$ is the gravitational acceleration at
the fluid interface (cf. Chandrasekhar 1961). Since the
dominant wavelength for the gas ablation by Kelvin-Helmholtz
instability is the order of $r_0$ (Murray et al. 1993), the
inequality (\ref{ce2}) for
$\rho_{\rm CG} \ll \rho_{\rm g, avg}$ is transformed as
\begin{eqnarray}
\rho_{\rm CG} v_{\rm gal}^2
	> \frac{GM_0 \rho_{\rm g, avg}}{2\pi r_0}.
\end{eqnarray}
Using this inequality, it is found that the diffuse gas
component in these galaxies is removed if their core mass
$M_0$ does not exceed a critical value given by
\begin{eqnarray}
M_{\rm KH}=1.60\times10^{12} \left( \frac{F}{0.1}
		\right)^{-\frac{7}{2}}
 \left( \frac{n_{\rm CG}}{10^{-4}{\rm cm}^{-3}}
		\right)^{\frac{7}{2}}
	\makebox[3em]{} \nonumber \\ \times
 \left( \frac{v_{\rm gal}}{10^{ 3}{\rm km ~s}^{-1}}
		\right)^{7} M_\odot.
\end{eqnarray}
This value corresponds to masses characteristic for a giant
galaxy which indicates that dwarf galaxies should in general
be affected by Kelvin-Helmholtz stripping. The gas ablation
due to Kelvin-Helmholtz instability is effective for massive
dwarf galaxies if the mass-loss time-scale is smaller than the
characteristic time-scale a galaxy spends spends within a
cluster environment.

According to Nulsen (1982) the mass-loss rate from the galaxy
through Kelvin-Helmholtz instability is estimated as
\begin{eqnarray}
\dot{M}_{\rm KH}=\pi r_0^2 \rho_{\rm CG} v_{\rm gal},
\end{eqnarray}
the gas removal time-scale of the dwarf galaxies is
therefore given by
\begin{eqnarray}
\tau_{\rm KH}&=&\frac{F M_0}{\dot{M}_{\rm KH}}, \\
&=&2.19\times10^9
 \left(\frac{F          }{0.1                     }
					\right)
 \left(\frac{M_0        }{10^{ 9} M_\odot         }
					\right)^\frac{1}{7}
 \left(\frac{n_{\rm CG}}{10^{-4}{\rm cm 	  }^{-3}}
					\right)^{-1}
					\nonumber \\
&& \makebox[7em]{} \times
 \left(\frac{v_{\rm gal}}{10^{ 3}{\rm km~s}^{-1}}
		\right)^{-1} ~{\rm yr}. \label{taukh}
\end{eqnarray}
Moreover, it should be noted that the characteristic
time-scale, normalized by the dynamical time
\begin{eqnarray}
\tau_{\rm dyn}
	&=&\sqrt{\frac{3 \pi}{32 G \rho_{\rm d, avg}}}, \\
	&=&8.93\times10^7\left(\frac{M_0}{10^9M_\odot}
			\right)^\frac{1}{7} {\rm yr},
\end{eqnarray}
where $\rho_{\rm d, avg}$ is the mean density of the CDM halo
inside $r_0$, is given by
\begin{eqnarray}
\frac{\tau_{\rm KH}}{\tau_{\rm dyn}}=24.5
  \left(\frac{F             }{0.1                     }\right)
  \left(\frac{n_{\rm CG}}{10^{-4}{\rm cm    }^{-3}}\right)^{-1}
	\makebox[7em]{} \nonumber \\  \times
  \left(\frac{v_{\rm gal}}{10^{ 3}{\rm km~s}^{-1}}\right)^{-1}.
\end{eqnarray}
This equation does not have an explicit dependence on the mass
of the CDM halo. Massive dwarf galaxies lose the
extended gas within $\sim 25 \tau_{\rm dyn}$ for $n_{\rm CG}=
10^{-4}$ cm$^{-3}$ and $v_{\rm gal}=1000$ km s$^{-1}$.
Consequently, we expect that even for the massive dwarf galaxy
the extended gas will be stripped through Kelvin-Helmholtz
instability within a short time-scale.

On the other hand, Balsara, Livio, \& O'Dea (1994)
demonstrated that the gas accretes from downstream into the
core in their hydrodynamical simulations of the interaction
between the ICM and the giant galaxy. Since the dwarf galaxies
have shallower potential wells than the giant galaxies, we
neglect the effect of the gas accretion form backward into the
galaxy. It may play however a role for the massive dwarf
galaxies where our estimation of the gas ablation might not be
suitable.

\section{Numerical Model}
In addition to the effect of the accretion inflow into the
core from downstream, the above analytic arguments neglect the
geometrical effects between the gas in dwarf galaxies and the
ICM and the presence of the Rayleigh-Taylor instability.
Moreover, the compressibility of the gas may become important
because the each galaxy moves in the cluster of galaxies with
the transonic velocity that is corresponding to the velocity
dispersion of the cluster of galaxies. Therefore, a realistic
analysis that uses a hydrodynamic simulation is necessary to
examine the effect of the gas removal due to the interaction
between a dwarf galaxy and an ICM. In this section, we
describe the numerical model of ram pressure stripping from
dwarf galaxies.

\subsection{Initial conditions and simulation method}
For modeling the interaction between a galaxy and an ICM,
Portnoy, Pistinner \& Shaviv (1993) showed that a treatment of
the protons and the electrons as two fluids resulted in
negligible differences for the mass of the gas inside the
galaxy. In this paper, therefore, the evolution of the gas is
described by the hydrodynamic equations for the single perfect
fluid. The continuity equation, the momentum equation, and the
thermal energy equation are given by
\begin{eqnarray}
\frac{\partial \rho}{\partial t}
	+ \nabla \cdot (\rho \v) = 0,
\end{eqnarray}
\begin{eqnarray}
\frac{\partial \rho \v{v}}{\partial t}
	+ \nabla \cdot (\rho \v \v)
	+ \nabla P =-\nabla \Phi_{\rm d},
\end{eqnarray}
and
\begin{eqnarray}
\frac{\partial \rho e}{\partial t}
	+ \nabla \cdot (\rho e \v)
	+ \nabla (P \v)
	=-\rho\v \cdot \nabla \Phi_{\rm d},
\end{eqnarray}
where $\rho$ is the gas density, \v is the gas velocity, $P$
is the gas pressure, $\gamma (=5/3)$ is the adiabatic index,
and $e$ is the total specific energy
\begin{eqnarray}
e = \frac{1}{2} v^2 + \frac{1}{\gamma-1} \frac{P}{\rho}.
\end{eqnarray}

Initial conditions are generated using the descriptions in
\S 2. We neglect the effects of the self-gravity of the gas,
radiative cooling, the heating by supernovae and stellar winds
from massive stars, and the photoionization by ultraviolet
radiation background. We discuss them in \S 5.
The equations are solved by a finite difference code VH-1.
VH-1 is based on the piecewise parabolic method (PPM)
described by Colella \& Woodward (1984) and was written and
tested by the numerical astrophysics group at the Virginia
Institute for Theoretical Astrophysics. Since the PPM scheme
has a great advantage due to  the reduction of numerical
viscosity, all fluid interfaces are sharply preserved and
small-scale features can be resolved. This scheme is,
therefore, suited for this class of problems.

The system is assumed to be axial symmetric and described
by cylindrical geometry. The center of the fixed
gravitational potential given by equation (\ref{pot}) is
located on the symmetric axis and the gas distribution
of the dwarf galaxy is set up using equation (\ref{rhog}).
The reflecting boundary condition is adopted at the bottom
boundary that corresponds to the symmetric axis and the ICM
flows continuously from the left stream boundary parallel to
the axis. The outflow boundary conditions, by imposing for
each variable a zero gradient ($d/dr=0$), are adopted at the
top and the right stream boundaries.
Unfortunately, this condition does not eliminate reflection
waves from the boundaries. Thus, the simulation box is
separated by two parts such as the inner rectangle $8r_0$
long and $3r_0$ wide and an outer surrounding part to keep
the boundaries far away.
The grids are equally spaced with 100 zones per core radius
for the higher resolution runs or 50 zones for the lower
resolution runs in the inner part, and are exponentially
spaced in the outer part.
A useful additional effect of this non-uniform grid is the
increased numerical dissipation of disturbances that propagate
to a large distance. This helps solving the problem of
residual reflection waves from the boundary.

\subsection{Results of simulations}
We have performed a parameter study varying the core mass
$M_0$ of the CDM halo from $10^6 M_\odot$ to $10^{10} M_\odot$,
varying the relative velocity of the dwarf galaxy
$v_{\rm gal}$ from 500 km s$^{-1}$ to 1000 km s$^{-1}$, and
varying the number density $n_{\rm CG}$ of the ICM from
$10^{-5}$ cm$^{-3}$ to $10^{-3}$ cm$^{-3}$. Table 1 displays
the model parameters of the ICM in this paper.

Figure 2 shows snapshots of the run for the model ($c$) of
$M_0=10^7M_{\odot}$, $n_{\rm CG}=10^{-4}$ cm$^{-3}$, and
$v_{\rm gal}=1000$ km s$^{-1}$ as a function of elapsed time
at $5.61\times10^6, 2.86\times10^7, 4.58\times10^7,
7.42\times10^7$, and $9.63\times10^7$ yr.
The left and the right colour images show the logarithmic
density distribution and the logarithmic pressure
distribution, respectively. Arrows in the right panels
indicate the velocity vector of the gas flow at each point
and their lengths are proportional to the absolute values of
their velocity. The CDM halo is located at the origin and the
ICM flows in from the left side.

The upper panel shows the early phase of the interaction
between the ICM and the gas of the dwarf galaxy. It is clearly
seen that the forward shock that propagates through the gas of
dwarf galaxy, and the reverse shock that propagates through
the ICM are formed in the downstream and the upstream of the
contact discontinuity respectively.
The propagation of the forward shock causes the strong
compression of the gas in the dwarf galaxy as seen in the
upper and the second panel. Below the second panel,
Kelvin-Helmholtz instabilities are observed at the contact
discontinuity.

Since the instabilities are suppressed by the gravity, the
growing modes are only the disturbances of the large wave
numbers that have the shortest growth time. Disturbances with
small wave numbers grow while moving away from the galaxy.
The third panel shows that the gas associated with the dwarf
galaxy is accelerated by the ram pressure of the ICM and is
pushed out of the potential well. Since the front of the
accelerated gas is highly Rayleigh-Taylor unstable, there are
many small irregularities in front of the contact
discontinuity. Accurately, the feature seen on the symmetric
axis ($R=0$) at $Z\sim0.5 r_0$ is caused by pure
Rayleigh-Taylor instability. The combination of
Rayleigh-Taylor instability and Kelvin-Helmholtz instability
forms other irregularities because the gas flows tangentially
to the fluid interface as seen in the third and right panel.
However, the perturbations do not grow significantly because
the gravity of the CDM halo suppresses Rayleigh-Taylor
instability. The gas removal time-scale nicely agrees with
the analytical estimate in equation (\ref{tauss}) which
predicts $\tau_{\rm IS}=5.42\times10^7$ yr.

Figure 3 shows snapshots of the run for the model ($c$) of
$M_0=10^{10}M_{\odot}$, $n_{\rm CG}=10^{-4}$ cm$^{-3}$, and
$v_{\rm gal}=1000$ km s$^{-1}$ as a function of elapsed time
at $6.84\times10^7, 6.99\times10^8, 1.15\times10^9,
2.18\times10^9$, and $3.26\times10^9$ yr.

The early evolution of the interaction between the gas and the
ICM is almost the same as in the case of $M_0=10^7M_\odot$.
However, since the gravitational potential is deep and the
central thermal pressure is larger than the ram pressure of
the ICM as shown in the analytical model, the mass-loss
process is not instantaneous but mild ablation due to
Kelvin-Helmholtz instabilities occur. Moreover, the gas
interface is Rayleigh-Taylor stable because the gas acceleration
is week. The velocity field in the right bottom panel reveals
that the flow turns back and an accretion inflow into the core
develops close to the symmetry axis on the downstream side.
Accretion occurs quasi-periodically as a radial-pumping mode.
The condition of the flow is summarized
as follows.
\begin{enumerate}
\item
This flow from the downstream side and the post-shock gas from
the upstream side cause  increased compression mainly along
the symmetry axis.
\item
The gas in the galaxy, therefore, acts by expanding sideways
as seen in the third panel. Since this expansion increases the
cross section of the interaction between the galactic gas and
the ICM, the mass-loss rate increases.
\item
The expansion decreases the pressure gradient around the
galaxy center and the gas contracts to the galaxy center by
the gravitational force. Then the system recovers the
quasi-static states again.
\item
This cycle repeats again.
\end{enumerate}
This process is effective in the simulations of a massive
dwarf galaxy or of a small ram pressure of the ICM. For this
specific process, the analytic condition given by \S3.2 fails
to describe the gas ablation from the dwarf galaxy.

Figure 4 shows the evolution of the gas mass inside the core
radius around the center of the dark matter halo as a function
of time for core masses of the CDM
halo from $10^6M_\odot$ to $10^{10}M_\odot$. Each curve is
categorized by the model parameters (see Table 1).
The gas is instantaneously stripped in low-mass dwarfs or for
large ram pressures. Due to the accretion inflow into the
galactic core from the down stream, the net mass-loss rate
becomes small. Moreover, we can observe oscillations of the
total mass inside a core radius due to the mass accretion.
These oscillation modes have periods of several dynamical
timescales. Even for the massive dwarf galaxy
($M_0=10^{10}M_\odot$), the gas could be instantaneously
stripped if the ram pressure is large as e.g. in the cores of
galactic clusters.

Figure 5 summarizes the results of our simulations. This
diagram shows the  mass-loss rate due to ram pressure
stripping as a sequence of the core mass.
Filled circles indicate the cases where the gas in the core
radius is completely stripped within 1 Gyr. In these cases the
ram pressure stripping by the ICM is very effective and the
whole gas in the potential well is rapidly removed.
Open circles indicate the cases where the gas in the core
radius is not completely stripped within 2 Gyr. In these cases,
the ram pressure stripping by the ICM is not so effective.
The thin line indicates the relation of
\begin{eqnarray}
\rho_{\rm CG} v_{\rm gal}^2= \frac{G M_0 \rho_{\rm g0}}{3r_0},
\end{eqnarray}
which is the instantaneous stripping condition described in
\S3.1. This line divides the parameter space into two regions.
Stripping is effective for the upper region and is
ineffective for the lower region. This relation roughly
agrees with the numerical experiments in the range from
$M_0=10^6M_\odot$ to $M_0=10^{10}M_\odot$. The dashed line
indicates the relation of
\begin{eqnarray}
\rho_{\rm CG} v_{\rm gal}^2= \frac{G M_0 \rho_{\rm g, avg}}
				{2\pi r_0},
\end{eqnarray}
which is the Kelvin-Helmholtz stripping condition described
in \S3.2. Stripping by Kelvin-Helmholtz instability is
effective for the upper region and is not effective for the
lower region. Since there is mass accretion from the back of
the galaxy, this relation is not well reproduced by the
numerical experiments.

Using the data of intermediate cases that are shown by filed
triangles in Figure 5, we find a critical core mass for
effective ram pressure stripping of
\begin{eqnarray}
M_{\rm cr}= 2.52\times10^9
\left( \frac{n_{\rm CG}}{10^{-4}{\rm cm}^{-3}}
			\right)^{\frac{5}{2}}
\left( \frac{v_{\rm gal}}{10^{ 3}{\rm km ~s}^{-1}}
			\right)^{5} M_\odot.
			\label{mcr}
\end{eqnarray}
The thick line shows the loci of this critical core mass.

In Figure 5, $r_{\rm CG}$ is the distance from the center of
galaxy cluster assuming the isothermal $\beta$-model
\begin{eqnarray}
n_{\rm CG}=n_{\rm CG0}
\left\{ 1+
	\left( \frac{r_{\rm CG}}{r_{\rm CG0}}
			\right)^2 \right\}
			^{-\frac{3}{2}\beta}, \label{beta}
\end{eqnarray}
with $\beta=0.6, n_{\rm CG0}=2\times10^{-3}$ cm$^{-3},
r_{\rm CG0}=0.25$ Mpc, and $\sigma_{\rm CG}=866$ km s$^{-1}$,
where $n_{\rm CG0}$ and $r_{\rm CG0}$ is the central number
density and the core radius of an ICM and $\sigma_{\rm CG}$ is
the line-of-sight velocity dispersion of the galaxy cluster
which corresponds to the relative velocity of the dwarf galaxy.
If the galaxy number density follows an approximate King model
as
\begin{eqnarray}
n_{\rm gal}\propto
\left\{1+\left( \frac{r_{\rm CG}}{r_{\rm CG0}}\right)^2
			\right\}^{-3/2},
\end{eqnarray}
the median radius of the galaxy distribution for
$r<10 r_{\rm CG0}$ is about $3.5 r_{\rm CG0}$.
The shaded region corresponds to ICM conditions inside this
radius. The diagram indicates that the process of the ram
pressure stripping is very effective for dwarf galaxies in the
clusters of galaxies. There, diffuse gas in dwarf galaxies
should rapidly be removed.

\section{Summary and discussion}
The physics of the interaction between the ICM and dwarf
galaxies confined by a surrounding CDM halo has been studied
in detail using analytical estimates and hydrodynamic
simulations. We have performed a parameter study varying the
core mass of the CDM, the relative velocity of the galaxy and
the density of the ICM. We find that the gas in dwarf galaxies
is rapidly removed in a typical cluster environment by
ram-pressure stripping.

Our results can be applied to clusters of galaxies, where
X-ray emission has been observed. Their gas number density
$n_{\rm CG}$ at the median radius of the galaxy distribution
and the line-of-sight velocity dispersion $\sigma_{\rm CG}$
are plotted in Figure 6. Data for the central gas density and
$\beta$ are from  Jones \& Forman (1999) and
Briel, Henry \& B\"{o}hringer (1992), and the velocity
dispersions are from the same sources including also
Hughes (1989).
One particular example, the Coma cluster with
$n_0=2.89\times10^{-3}$ cm$^{-3}, \beta=0.75$ and
$\sigma_{\rm CG}=1010$ km s$^{-1}$, is indicated by a filled
circle. Dotted lines are loci of constant $M_{\rm cr}$ given
by equation (\ref{mcr}) ranging from $10^7M_\odot$ to
$10^{11}M_\odot$. The lines with their various core masses
and maximum circular velocities are shown. They indicate that
galaxies should rapidly loose all of their diffuse gas by ram
pressure stripping. This figure shows that the gas in dwarf
galaxies is completely stripped in rich clusters of
galaxies within the short time-scale.

The effect of radiative cooling has been neglected in our
simulations. we estimate the role of radiative cooling in our
models here. Since for low-mass dwarf galaxies
($M_0 \lesssim 10^9M_\odot$), the ram pressure already exceeds the
gravitational force, radiative processes will not be able to
affect the gas stripping. However, it might play a role for
massive dwarf galaxies ($M_0 \gtrsim 10^9M_\odot$) which have
deeper potential wells. The cooling time-scale is defined as
(Efstathiou 1992)
\begin{eqnarray}
\tau_{\rm cool}=\frac{3}{2}\frac{1}{\mu^2(1-Y)^2}
	\frac{k_{\rm B}T}{n_{\rm g}\Lambda},
\end{eqnarray}
where $Y(=0.25)$ is the helium mass fraction, $n_{\rm g}$ is
the gas number density, and $\Lambda$ is cooling rate. If we
evaluate this time-scale for our standard model
$M_0=10^9M_\odot$ using conditions immediately behind the
leading shock, we find that $\tau_{\rm cool}=3.18\times10^9$
yr assuming $T=10^7$ K,
$\Lambda=10^{-23}$ erg s$^{-1}$ cm$^{-3}$, and
$n_{\rm g}=1.0$ cm$^{-3}$ from the result of our simulations.
This time-scale is larger than the instantaneous stripping
time-scale ($\tau_{\rm IS}=2.03\times10^8$ yr) that is given
by equation (\ref{tauss}). This implies that gas stripping
from the massive dwarf galaxies is affected by the radiative
cooling.

However, we would then also need to consider the
effects of the star formation and subsequent feedback process
such as stellar winds and supernovae heating from massive
stars. These feedback processes supply thermal energy to the
gas and prevent efficient cooling (Dekel \& Silk 1986;
Mori et al. 1997; Mori et al. 1999).
In addition, the photoionization due to the ultraviolet
background radiation prevents gas cooling and keeps the gas
hot (Efstathiou 1992; Babul \& Rees 1992).
Furthermore, though we studied only the interaction
between the ICM and a hot interstellar medium in this paper,
it is very interesting to examine also the fate of the cool 
and dense interstellar medium in a dwarf galaxy.
In this case, the stripping history of the gas may be quite
different even for the less-massive dwarf
galaxies ($M<10^9 M_\odot$).
In a series of forthcoming studies, we will report the
results of taking into account the multi-phase states of
the gas with cool components in a galaxy, including the
effect of the radiative cooling of the gas, energy input
from stars, and the ultraviolet background radiation.

We conclude that the ram pressure stripping of the diffuse
gas from dwarf galaxies by the ICM is of prime importance
to understand the morphology-density relation for dwarf
galaxies and the chemical evolution of dwarf galaxies.
Most of the heavy elements ejected by massive stars through
Type II supernovae are found in the warm and hot diffuse gas.
In order for these heavy elements to be incorporated into
stars, the hot gas has to cool and form cold clouds again.
Our models indicate that this will be impossible for dwarf
galaxies in clusters of galaxies, as the gas is stripped
before condensing into clouds. Ram pressure stripping
therefore plays a significant role for the chemical evolution
of dwarf galaxies. Our results imply that most of the
chemical evolution of dwarf galaxies must have occured at high
redshift, before the virialized cluster had formed.

%
%

\begin{acknowledgements}
This work has been supported in part by the Research 
Fellowship of the Japan Society for the Promotion of Science 
for Young Scientists (6867) of the Ministry of Education, 
Science, Sports, and Culture in Japan. 
M.M. would like to thank the Max-Planck-Institut f\"{u}r 
Astronomie for its hospitality throughout this research.
Numerical calculations were carried out at Max-Planck-Institut 
f\"{u}r Astronomie in Germany and at the Astronomical Data 
Analysis Center of the National Astronomical Observatory in 
Japan.
\end{acknowledgements}

%
%

%
%

\clearpage

\figcaption[f1.ps]{
The density (upper) and the circular velocity (lower) of the
Burkert's (1995) profile. Each curve is divided by the core
mass of the dark matter halo which ranges from $10^6M_\odot$
(left) to $10^{10}M_\odot$ (right).
}

\figcaption[f2.ps]{
Snapshots for the model with core mass $M_0=10^7M_{\odot}$,
number density of the ICM $n_{\rm CG}=10^{-4}$ cm$^{-3}$, and
relative velocity of the galaxy $v_{\rm gal}=1000$ km s$^{-1}$
as a function of elapsed time at $5.61\times10^6$,
$2.86\times10^7$, $4.58\times10^7$, $7.42\times10^7$, and
$9.63\times10^7$ yr. The left and the right colour images show
the logarithmic density distribution and the logarithmic
pressure distribution, respectively. Arrows in the right
panels indicate the velocity vector of the gas flow at each
point.
}

\figcaption[f3.ps]{
Snapshots for the model with core mass $M_0=10^{10}M_{\odot}$,
number density of the ICM $n_{\rm CG}=10^{-4}$ cm$^{-3}$, and
relative velocity of the galaxy $v_{\rm gal}=1000$ km s$^{-1}$
as a function of elapsed time at $6.84\times10^7$,
$6.99\times10^8$, $1.15\times10^9$, $2.18\times10^9$, and
$3.26\times10^9$ yr. The left and the right colour images show
the logarithmic density distribution and the logarithmic
pressure distribution, respectively. Arrows in the right
panels indicate the velocity vector of the gas flow at each
point.
}

\figcaption[f4.ps]{
The evolution of the gas mass inside the core radius of the
gravitational potential of the CDM halo as a function of time.
Each column corresponds to a core mass of the CDM halo which
varies from $10^6M_\odot$ to $10^{10}M_\odot$. Each curve is
categorized by the model parameters(dash-dot-dot-doted line:
$~n_{\rm CG}=10^{-3}$ cm$^{-3}$ and
$v_{\rm gal}=1000$ km s$^{-1}$, doted line:
$~n_{\rm CG}=10^{-3}$cm$^{-3}$ and
$v_{\rm gal}=500 $ km s$^{-1}$,
dash-doted line:$~n_{\rm CG}=10^{-4}$ cm$^{-3}$ and
$v_{\rm gal}=1000$ km s$^{-1}$, dashed line:
$~n_{\rm CG}=10^{-4}$cm$^{-3}$ and
$v_{\rm gal}=500 $ km s$^{-1}$, and solid line:
$~n_{\rm CG}=10^{-5}$ cm$^{-3}$ and
$v_{\rm gal}=1000 $ km s$^{-1}$). Each holizontal line
indicates the time scale of $10\tau_{\rm dyn}$.
}

\figcaption[f5.ps]{
The mass loss time-scale for dwarf galaxies as a result of
ram pressure stripping $\rho_{\rm CG} v_{\rm gal}^2$ due to
the ICM as a sequence of the core mass $M_0$ and the maximum
circular velocity $v_{\rm c, max}$ of dwarf galaxies.
Filled circles indicate the case that the gas is stripped
within 1 Gyr. Open circles show that the gas is not stripped
within 2 Gyr. Intermediate cases are marked by filed
triangles. The shaded region corresponds to the inside of the
median radius of the galaxy distribution for a typical galaxy
cluster fitted by $\beta$-model
$n_{\rm CG}=n_{\rm CG0} \{ 1+ (r_{\rm CG}/r_{\rm CG0})^2
\}^{-3\beta/2}$, with $\beta=0.6$, the central number density
$n_{\rm CG0}=2\times10^{-3}$ cm$^{-3}$, the core radius
$r_{\rm CG0}=0.25$ Mpc, and the line-of-sight velocity
dispersion $\sigma_{\rm CG}=866$ km s$^{-1}$. $r_{\rm CG}$ is
the distance from the center of cluster.
}

\figcaption[f6.ps]{
A scaled gas density is plotted against velocity dispersion
for clusters of galaxies. Assuming a distribution of the gas
that follows the $\beta$-model, the gas number density is
given at the median radius of the galaxy distribution. Data
for the central gas density and $\beta$ are from Jones \&
Forman (1999) and Briel, Henry \& B\"{o}hringer (1992), and
the velocity dispersions are from these sources and from
Hughes (1989). The filled circle corresponds to the Coma
cluster. Solid lines are loci of constant $M_{\rm cr}$ which
is the critical core mass of the CDM halo, and the lines are
labeled with their core mass and the maximum circular
velocity. The lines indicate that dwarf galaxies should lose
instantaneously all of their gas by ram pressure stripping.
}

%
%


\clearpage

\begin{references}
\reference{} Babul, A., \& Rees, M. 1992, MNRAS, 255, 346
\reference{} Balsara, D., Livio, M., \& O'Dea, C.P. 1994,
			ApJ, 437, 83
\reference{} Binggeli, B., Tammann, G., \& Sandage, A. 1987,
			ApJ, 94, 251
\reference{} Binggeli, B., Tarenghi, M., \& Sandage, A. 1990,
			A\&A, 228, 42
\reference{} B\"{o}hringer, H., Nulsen, P.E., Braun, R., \&
			 Fabian, A.C. 1995, MNRAS, 274, L67
\reference{} Briel, U.G., Henry, J.P., \& B\"{o}hringer, H.
			1992, A\&A, 259, L31
\reference{} Burkert, A. 1995, ApJ, 447, L25
\reference{} Burkert, A., \& Ruiz-Lapuente, P.  1997, ApJ, 480, 297
\reference{} Chandrasekhar, S. 1961, {\it Hydrodynamic and
			Hydromagnetic Stability}
			(Oxford:Oxford University Press)
\reference{} Cole, S., Arag\'{o}n-Salamanca, A., Frenk, C.S.,
			Navarro, J.F., \& Zepf, S.E.,
			1994, MNRAS, 271, 781
\reference{} Colella, P., \& Woodward, P. 1984, JCP, 54, 174
\reference{} Dekel, A., \& Silk, J. 1986, ApJ, 303, 39
\reference{} Dressler, A. 1980, ApJ, 236, 351
\reference{} Dubinski, J., \&  Carlberg, R. 1991, ApJ, 378, 496
\reference{} Efstathiou, G. 1992, MNRAS, 256, 43P
\reference{} Ferguson, H., \& Sandage, A. 1988, AJ, 96, 1520
\reference{} Flores, R.A., \& Primack, J.R. 1994, ApJ, 427, L1
\reference{} Fukushige, T., \& Makino, J. 1997, ApJ, 477, L9
\reference{} Hughes, J.P. 1989, ApJ, 337, 21
\reference{} Ikeuchi, S., 1986, Ap\&SS, 118, 509
\reference{} Irwin, J.A., Seaquist, E.R., Taylor, A.R., \&
			Duric, N. 1987, ApJ, 313, L91
\reference{} Jones, C., \& Forman, C. 1999, ApJ, 511, 65
\reference{} Katz, N., Weinberg, D.H., \& Hernquist, L., 1996,
			ApJS, 105, 19
\reference{} Moore, B. 1994, Nature, 370, 629
\reference{} Moore, B., Governato, F., Quinn, T., Stadel, J.,
			 \& Lake, G. 1998, ApJ, 499, L5
\reference{} Mori, M., Yoshii, Y., \& Nomoto, K. 1999,
			ApJ, 511, 585
\reference{} Mori, M., Yoshii, Y., Tsujimoto, T., \& Nomoto, K.
			1997, ApJ, 478, L21
\reference{} Murray, S.D., White, S.D.M., Blondin, J.M., \&
			Lin, D.N. 1993, ApJ, 407, 588
\reference{} Navarro, J.F., Frenk, C.S., \& White, S.D.M. 1996,
			ApJ, 462, 563
\reference{} Navarro, J.F. \& Steinmetz, M., 1997, ApJ, 478, 13
\reference{} Nulsen, P.E.J. 1982, MNRAS, 198, 1007
\reference{} Portnoy, D., Pistinner, S., \& Shaviv, G. 1993,
			ApJS, 86, 95
\reference{} Rees, M.J., 1986, MNRAS, 218, 25p
\reference{} Thuan, T.X., Alimi, J.M., Gott, J.R., \&
			Schneider, S.E. 1991, ApJ, 370, 25
\reference{} White, D.A., Fabian, A.C., Forman, W., Jones, C,
			 \& Stern, C. 1991, ApJ, 375, 35
\reference{} White, S.D.M. \& Frenk, C.S., ApJ, 379, 52
\reference{} Yoshii, Y., \& Arimoto, N., 1987, A\&A, 188, 13
\end{references}
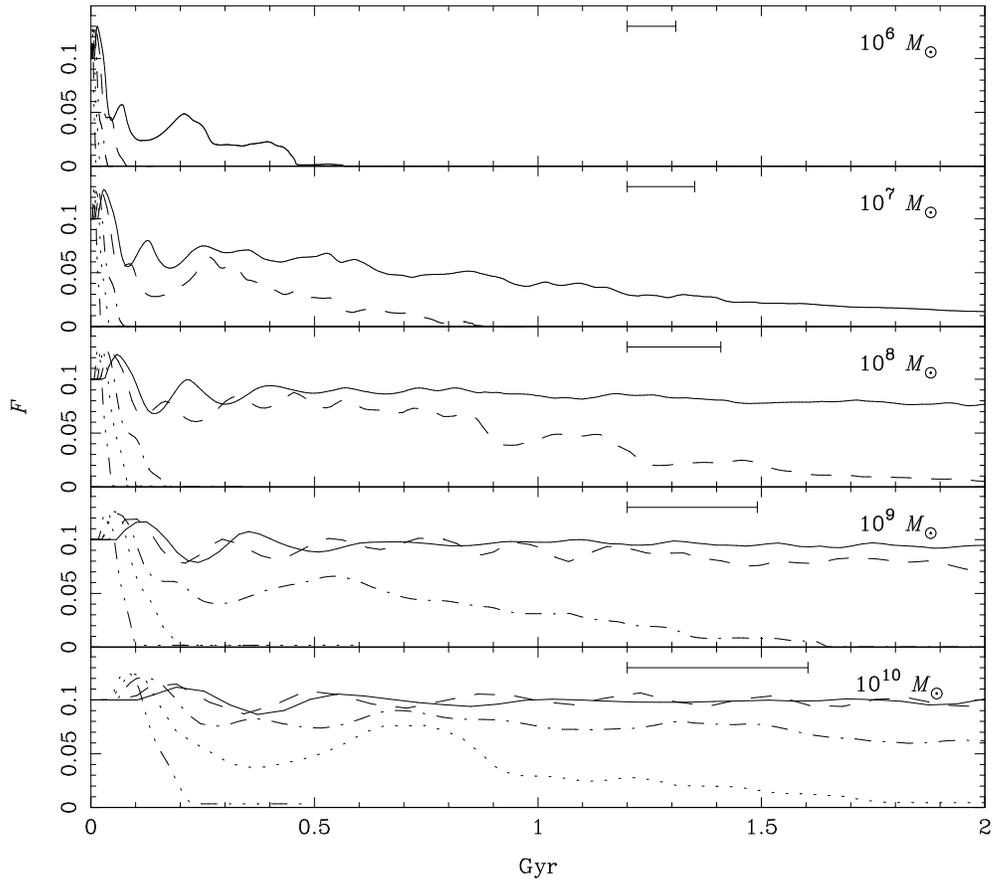
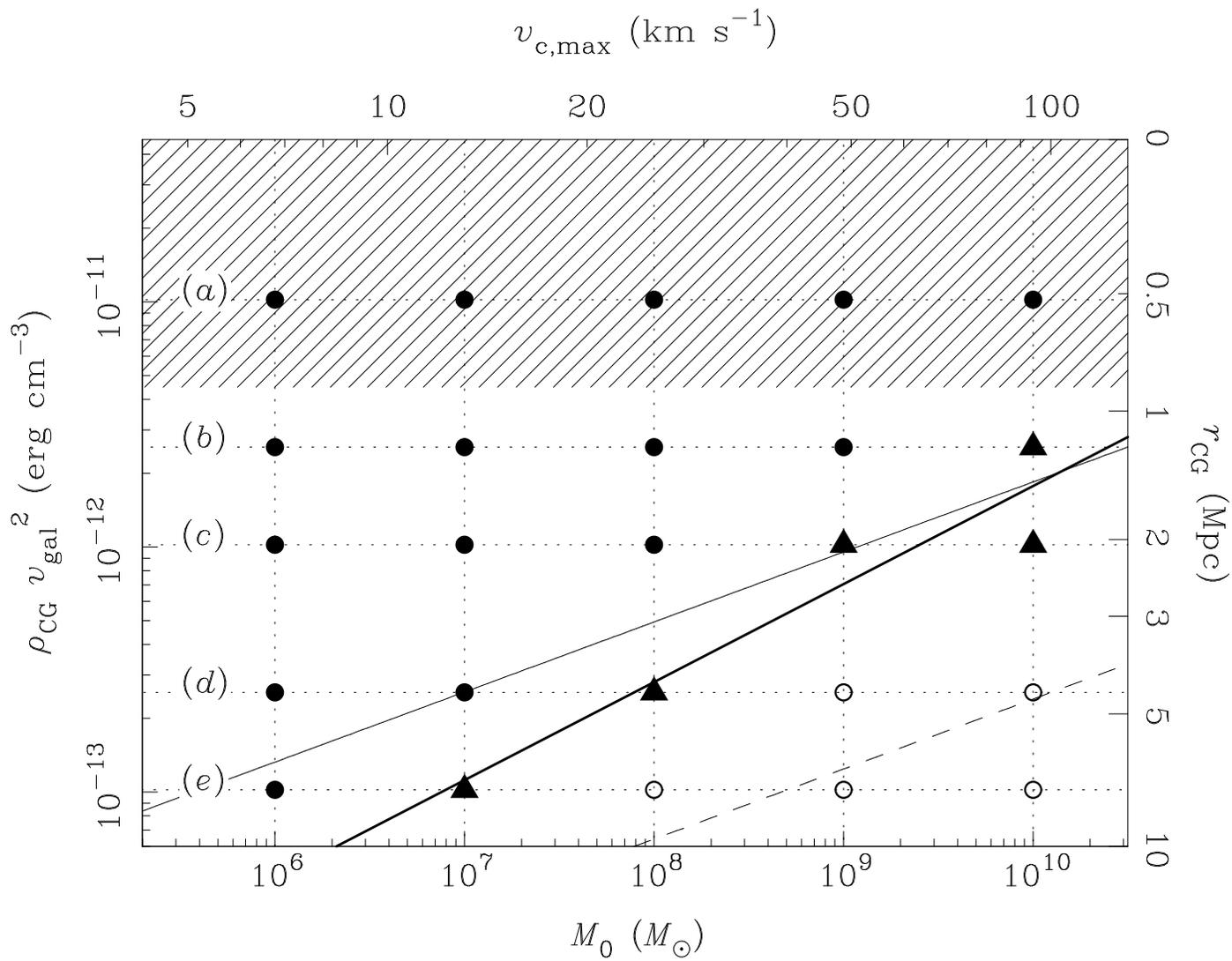
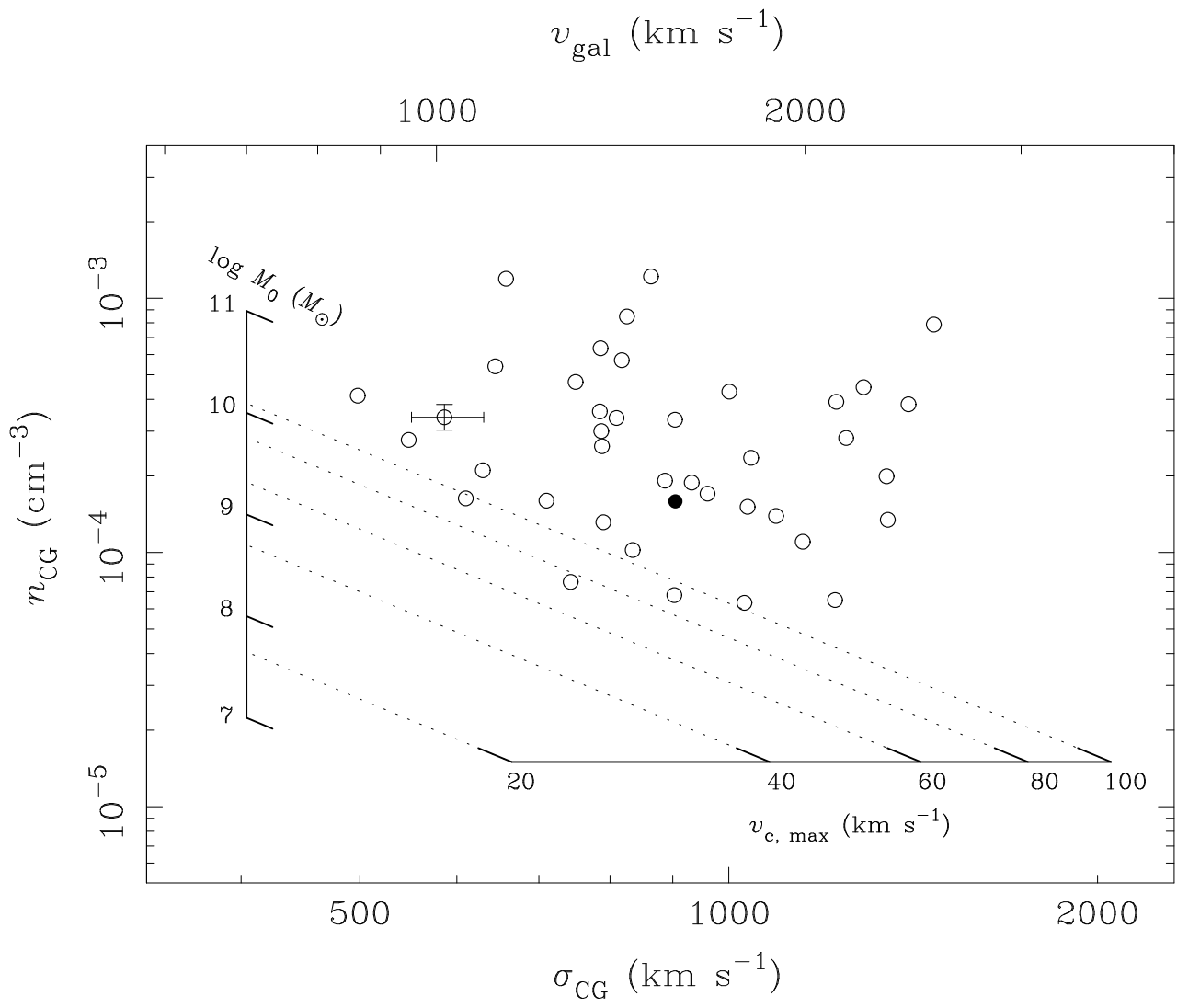
\end{document}